\documentclass[review]{elsarticle}

\usepackage{lineno,hyperref}
\modulolinenumbers[5]

\usepackage{graphicx}
\usepackage{latexsym}
\usepackage{xkeyval}
\usepackage{nccmath}
\usepackage{tikz}\usetikzlibrary{patterns}
\usepackage{amsmath}
\usepackage{amsfonts}
\usepackage{xfrac}
\usepackage{subcaption}
\usepackage{bm}
\usepackage{multirow}
\newcommand\numberthis{\addtocounter{equation}{1}\tag{\theequation}} 

\allowdisplaybreaks
\DeclareMathOperator*{\argmax}{arg\,max}

\usepackage{xcolor}

\begin{document}

\begin{frontmatter}

\title{Predicting Voting Outcomes in the Presence of Communities, Echo Chambers and Multiple Parties}


\author[warwickaddress]{Jacques Bara}

\author[omeraddress]{Omer Lev}

\author[warwickaddress]{Paolo Turrini\corref{mycorrespondingauthor}}
\cortext[mycorrespondingauthor]{Corresponding author}
\ead{p.turrini@warwick.ac.uk}

\address[warwickaddress]{University of Warwick, Coventry CV47AL, United Kingdom}
\address[omeraddress]{Ben-Gurion University, Beer-Sheva 8410501, Israel}

\begin{abstract}

When individuals interact in a social network their opinions can change, at times quite significantly, as a result of social influence.

In elections, for example, while they might initially support one candidate, what their friends say may lead them to support another. 
But how do opinions settle in a social network, as a result of social influence?

A recently proposed graph-theoretic metric, the {\em influence gap}, has shown to be a reliable predictor of the effect of social influence in two-party elections, albeit only tested on regular and scale-free graphs. Here, we investigate whether the influence gap is able to predict the outcome of multi-party elections on networks exhibiting community structure, i.e., made of highly interconnected components, and therefore more resembling of real-world interaction. 
To encode communities we build on the classical model of caveman graphs, which we extend to a richer graph family that displays different levels of {\em homophily}, i.e., how much connections and opinions are intertwined. 

Our contribution is three-fold. 
First, we study the predictive power of the influence gap in the presence of communities. We show that when there is no clear initial majority the influence gap is not a good predictor of the election outcome. When we instead allow for varying majorities, although the influence gap improves as a predictor, counting the initial partisan majority does consistently better, across all levels of homophily. Second, we study the combined effect of the more predictive metrics, as function of the homophily levels.
Using regression models, we demonstrate that the influence gap combined with the initial votes count does increase the overall predictive power for some levels of homophily. Third, we study elections with more than two parties. Specifically, we extend the definition of the influence gap to any number of parties, considering various generalisations, and show that the initial votes count has an even higher predictive power when compared to influence gap than it did in the two-party case.
\end{abstract}

\begin{keyword}
Social Networks\sep 
Opinion Dynamics\sep 
Voting \sep 
Communities \sep
Echo Chambers \sep
Influence Gap

\end{keyword}

\end{frontmatter}


\section{Introduction}
The role of social networks in shaping collective decisions is widely recognised in today's society. Recent political developments, from fake news to data leaks, have also brought to light their powerful role in potentially altering election results, and, therefore, their vulnerability to various forms of manipulation ~\cite{JWF15,GDGM15,SCVM18,BLSS18}.


As a consequence of the growing importance of social networks in our public discourse, the research on collective decision-making is now looking at the spread of information as a key factor in determining voting outcomes \cite{acemoglu-spread,Wilder-Vorobeychik2018,Federico-et-al-2019}, with important ramifications for network engineering \cite{Castiglioni_Ferraioli_Gatti_2020}. The multi-agent systems community, in particular, explored the algorithmic side of social dynamics in existing voting models (e.g., \citet{Tsang-Larson-2016}), studying novel possibilities to manipulate collective outcomes~\cite{Alon-Feldman-Omer-2015,baumeister2020manipulation, GrandiT16,BredereckE17} and making the study of social choice on social networks an active research enterprise~\cite{Grandi2017}. 

When studying the effect of social influence on collective decisions, establishing predictive metrics is paramount. 
A recent {\em Nature} paper by \citet{Stewart2019} has shown how a simple graph-theoretic metric, the {\em influence gap}, is highly predictive of how influence dynamics will impact the result of an election, turning minority views (with well-placed supporters) into strong majorities. 

More specifically, through computational simulations of a voter model, backed by social network experiments with human subjects, \citet{Stewart2019} found strong correlations between the outcome of the voters' decisions and their proposed metric. These results suggest that an increased presence in a voters' social neighbourhood (what they call {\em influence assortment}) is a good predictor of a party's chances to win elections. In other words, when voters update their preferences looking at their connections, it is the strategic positioning of a party's electorate that matters, rather than the initial majority (what they call {\em information gerrymandering}). Undoubtedly, a metric that allows us to forego the equilibrium computation of a highly complex system is an important practical tool, significantly simplifying the analysis of the opinion diffusion dynamics, a notoriously complex problem \cite{ChristoffG17, AulettaFG20, ChistikovEtAl20}. Moreover, it allows for a further understanding of the effects of manipulation, for example through the strategical placements of bots or zealots to alter the network dynamics.

The results in \citet{Stewart2019} are, however, based on a number of limiting assumptions, notably the analysis is carried out on two-party elections on regular graphs of degree 3 and large scale-free graphs. 

The main goal of this paper is to examine whether the metric's usage can be expanded beyond this fairly narrow family of graphs to more realistic-looking graphs and into the multi-party case. Our focus is on graphs which are characterised by the presence of a community structure~\cite{Girvan7821,Flake-Lawrence-2002}, allowing for phenomena such as echo chambers and {\em homophily}~\cite{Bakshy1130,Tsang-Larson-2016}. These are well-established patterns of real-world social networks, which can be observed, for example, in how Americans are sorting themselves into partisan communities~\cite{Bis09}, and should, in our view, be accounted for by any reasonable model of how social influence affects collective decisions. 

\paragraph{Our Contribution} We analyse multi-party elections on graphs with community structure, exploring the predictive power of the influence gap and comparing it against various other metrics. We first model communities using what we call {\em homophilic relaxed-caveman} graphs, which build on the classical clique-like community model of caveman graphs~\cite{Watts_1999}. The homophilic relaxed-caveman graphs introduce variance and reality-resembling interactions by determining connections as a noisy function of the degree of homophily. The key difference from \citet{Stewart2019} is that we look at networks with community structure, while they only consider regular and scale-free graphs. We compute how homophily and rewiring interact in these graphs to change the influence gap (Figure \ref{fig:hrc-surf-precise}) and we then examine the influence gap as a predictor of the final voting outcome, using ~\citet{Stewart2019}'s empirically backed opinion diffusion model and parameters.

We see that with equal initial representation but varying influence gap, the latter no longer correlates with the final voting outcome as it did in~\citet{Stewart2019} (Figure \ref{fig:equal-representation}). Once we look at settings which include varying initial partisan majorities, the gap generally correlates well with the final voting outcome (Figure \ref{fig:ig+maj_vs}). However, simply counting the initial majorities is an even better predictor (Figure \ref{fig:ig+maj_vs}) and remains consistently so for different levels of homophily and rewiring (Figure \ref{fig:pcc_lines}).  Using regression models, we then determine the interrelation between influence gap and initial majority to strengthen the predictions (Table \ref{table:lin_reg}).

Finally, we move to the study of the influence gap in elections with more than two parties. We look at multiple extensions of the originally proposed definitions, analysing their behaviour on simple graphs such as rings, cycles and stars. Using the extended setup we provide a theoretical analysis on clique-like communities and we observe empirically how the initial votes count has an even stronger predictive power when compared to influence gap than it did in the two-party case (Figure \ref{fig:multiparty_pcc_lines}).

\paragraph{Other Related Literature}
Discussion of how opinions and ideas spread in society flourished as a research field since Rogers's seminal work \cite{Rog62}, who introduced many of the concepts still underlying the field. Research then expanded to cases where agents have limited information~\cite{FS85,KS85}, including on graph structures~\cite{Blu93,LW08}. There was a particular focus on ``information cascades'' or ``herd mentality'', where choices are made sequentially, both when there is a ground truth~\cite{BHW92,MNT14,FILW14} and where there is none~\cite{ABKLT12,Wat02}. We use this basic assumption that people wish to conform to their surroundings in this paper, as well.

A closely related avenue of research concerns opinion diffusion models, where agents are recipients of social influence and opinions spread in a network. Research on this has been both empirical~\cite{BR87} and theoretical~\cite{Gra73,GLM01,Alon-Feldman-Omer-2015} (see overviews \cite{MMB90,You09}), including attempts to find influential nodes in the social graph~\cite{KKT03}. Computational models of opinion diffusion have looked at the fixed-point properties of the graph dynamics, in connection with consensus formation \cite{AulettaFG20} and its complexity \cite{ChistikovEtAl20}. An important stream of research has looked at how to control opinion diffusion by external intervention, for example through bribery \cite{BredereckE17}, false-name attacks \cite{BrillCFS16} or information control \cite{baumeister2020manipulation}, and we see our results as introducing effective heuristics for outcome prediction in those frameworks.

\paragraph{Paper Structure}
In Section \ref{sec:preliminaries} we introduce our setup and the basic graph-theoretic terminology, in particular the model of caveman-graphs and some basic observations on the influence gap. In Section \ref{sec:RC} we present our homophilic extension, together with the algorithms to control the homophily level and the rewiring probability. Section \ref{sec:dynamics} focuses on two-party elections, introducing the opinion diffusion dynamics and measuring the predictive power of the influence gap and other key metrics. These results are further discussed in Subsection \ref{sec:comb_met}, where we combine metrics and establish the effects of homophily and rewiring. In Section \ref{sec:multiparty} we delve into the multi-party case. We first consider potential extensions of the influence gap, studying them on simple graphs and then using computer-aided simulations and larger graphs in Section \ref{sec:3-party-large} to observe their predictive power in a setting with three parties. We conclude in Section \ref{sec:discussion} presenting various follow-up research directions.

\section{Influence Gap and Communities}\label{sec:preliminaries}
Consider an election with parties $\mathcal{P}$, where for each party $P\in\mathcal{P}$ there are $N_P>0$ supporters (voters). The $N$ total voters are placed on an undirected graph $\mathcal{G}=(V,E)$, where they are represented as nodes in the node set $V = \{1,2,\cdots, N\}$ while social connections are given by edges in the edge set $E \subseteq V\times V$. 

Let $p:V\to \mathcal{P}$ be the party assignment, representing individuals' initial opinions, such that, for example, $p(n)$ is the party of voter $n$. We can also apply this to a subset of nodes $U\subseteq V$, and denote by $p(U)$ the parties that all nodes in the subset $U$ vote for, i.e., $p(U) = \{p(n) \mid n\in U\}$. We denote the neighbourhood of $n\in V$, i.e., the set of social connections of voter $n$, by $\mathcal{N}_n = \{m \mid (n,m)\in E\}$. We also define the poll of a node $n$ as themselves with their neighbourhood, $\mathcal{N}'_n = n\cup\mathcal{N}_n$. For a node $n\in V$, denote by $\Delta^P_n$ be the fraction of $n$'s poll that votes for party $P$ and let $\bm{\Delta}_n = (\Delta^P_n \mid P \in \mathcal{P})$ be the vector of such polls; for brevity we shall often use $\Delta_n$ instead of $\Delta_n^{p(n)}$ when no ambiguity arises.

\subsection{Influence Gap}
Like much of the research on opinion dynamics (including, in particular, \citet{Stewart2019}), we shall, at least at first, focus on the two-party (or two-opinion) setting and we shall henceforth refer to the parties using two colours, red and blue. In other words, we fix a partisan structure $\mathcal{P} = \{R,B\}$ where the initial number of voters for the red (blue) party is $N_R$ ($N_B$), noticing that with only two parties it suffices to consider only the fraction of a node's poll that votes for their own party, i.e., $\Delta_n$. 
The multiparty generalisation ($|\mathcal{P}| \geq 2$) has non-trivial ramifications for a number of concepts, and will be investigated in detail from Section \ref{sec:multiparty} onwards.

In the two-party case, {\em influence assortment} \cite{Stewart2019} is, intuitively, the relative advantage of a party against their rival, and acts on two different levels: on the level of a single node $n$, denoted by $a_n$; and on the level of a party $P$ denoted by $A_{P}$. The {\em influence gap} (IG), which we denote by $G_P$, is the average advantage in influence assortment of party $P \in \{R,B\}$ (resp., $G_{P'}$ denotes its dual for party $P'\neq P$). Below are the formal definitions (as per \citet{Stewart2019}); note the use of the Kroneker delta $\delta(i,j)$, which is 1 if $i=j$ and 0 otherwise.

 \begin{ceqn}
 \begin{align}
 a_n &= \begin{cases} 
 \Delta_n &\Delta_n \geq \frac{1}{2} \\
 -(1-\Delta_n) &\Delta_n < \frac{1}{2} \\
 \end{cases} \label{eq:node_inf}\\
 A_P &= \frac{1}{N_P}\sum_{n \in V} a_{n}\delta(p(n),P) \label{eq:party_inf}\\
 G_P &= A_P - A_{P'} \label{eq:IG} 
 \end{align} 
 \end{ceqn}
Influence assortment on the level of nodes, $a_n$, can be thought of as the extent to which an agent's party is present in their own poll, and thus how much they can be influenced to vote for a different party. Its value (regardless of sign) highlights how homogeneous a node's neighbourhood is, while its sign indicates if a node belongs to the majority party in the local neighbourhood. The mean of node assortments over nodes of a single party is then the party assortment, $A_P$. The influence gap, $G_P$ can therefore be understood as the difference in how ``strategically placed'' a party is -- how much do its supporters interact with other parties (and therefore, able to be influenced by them). 

Throughout the paper we focus on the case of a strong party assignment (SPA), following \citet{Alon-Feldman-Omer-2015}, where each party is assigned a fixed fraction of nodes -- initially this will be a half but in later sections we consider non-equal representation. Weak party assignment (WPA), in contrast, assigns to every node a party $P$ with some probability, e.g., they are red with probability $\frac{3}{4}$. While we leave the treatment of WPA for future research, we note that our random graph generation models with non-equal representation are de facto working with a constrained form of WPA.

%


\subsection{Caveman Graphs}\label{prelimCaveman}

A {\em caveman graph} $\mathcal{G} = (V,E)$ is a set of $l$ isolated cliques each of size $k$~\cite{Watts_1999}. These  graphs encode a very basic form of community-structure without showing interesting variety or empirical relevance \cite{emp-study-cliques}\footnote{A connected version is formed by rewiring a single edge per clique it to a node in an adjacent clique along a central cycle~\cite{Watts_1999}, also known as a caveman graph. For the purposes of this paper a caveman graph will be taken to mean the unconnected set of isolated cliques.}, but we shall use them for the most basic theoretical insights and will build on them to develop more realistic-looking structures. 




\subsubsection{Influence gap in caveman graphs}
In caveman graphs the influence gap shows some interesting features in connection with the partisan majorities in each clique. To see why this is the case, consider a graph of $l$ cliques (labelled $1,\cdots,l$), each of size $k$. Let us take, without loss of generality, the perspective of the red party first. A red node $n$ in clique $c\in\{1,\cdots,l\}$, with total number of red nodes $x_c$, sees exactly $\Delta_n = \frac{x_c}{k}$ red nodes in their poll, as, by definition, they see all of the nodes inside their clique. This clearly holds for all other red nodes inside $c$, as well. Thus, the influence assortment of any red node in $c$ is equal, so that $a_n = a_m$ for all nodes $n,m\in c$; the sum of influence assortments over all red nodes in $c$ is therefore $x_c a_n$. 

Once all $x_c$ are known, we can group and loosely order cliques together into three types: those in which red holds a strict majority, those in which there's an exact tie and those in which blue holds a strict majority. To this end $\exists M, M' \in \{1,\cdots,l\}$ with $M \leq M'$ such that for $n\in V$:

\begin{ceqn}
\begin{equation*}
 x_c \begin{cases} > \sfrac{1}{2} &\text{ for } c \leq M\\
 = \sfrac{1}{2} &\text{ for } M < c \leq M'\\
 < \sfrac{1}{2} &\text{ for } M' < c \\
 \end{cases}
\end{equation*}
\end{ceqn}


In other words, $M$ is the number of cliques in which the red party hold a strict majority, $M'$ is the number of cliques where they hold a weak majority in and $\eta\equiv M'-M$ is the number of {\em marginal} cliques. Finally, denoting the sum of $x_c$'s from $c=1$ up to $c=d$ (for $1\leq d\leq l$) as $X_d \equiv \sum_{c=1}^d x_c$, knowing the cliques with red majority $M'$ allows us to calculate the red assortment on the level of the party, $A_R$.
\begin{ceqn}
\begin{align*}
 A_R &= \frac{1}{N_R}\sum_{n \in V} a_n \delta(p(n),R) \\
 &= \frac{1}{N_R} \Bigg( \sum_{c=1}^{M'} \big(x_c\frac{x_c}{k}\big) + \sum_{c=M'+1}^l x_c\big(\frac{x_c}{k}-1\big) \Bigg) \\
 &= \frac{1}{N_R} \Bigg( \sum_{c=1}^l \frac{x_c^2}{k}-(N_R-X_{M'}) \Bigg) \\
 A_R &= \frac{1}{N_R} \Bigg( \sum_{c=1}^l \frac{x_c^2}{k} + X_{M'} \Bigg)-1 \numberthis
\end{align*}
\end{ceqn}


We can find the equivalent for the blue party by making two observations. First, the influence assortment of any blue node $n\in c$ is $b_n = -a_m$ for a red node $m\in c$. Second, the number of blue nodes in clique $c$ is $y_c = k-x_c$ hence the number of blue nodes up to clique $d$ is $Y_d \equiv \sum_{c=1}^d y_c = dk-X_d$. We note that since there are $M$ cliques that contain a strict red majority, then equivalently in these $M$ cliques blue is strictly a minority. Following similarly from the red party, the influence assortment of the blue party, $A_B$, in terms of red counts $x_n$ is thus as follows.

\begin{align*}
 &A_B= \frac{1}{N_B} \Bigg( \sum_{c=1}^l \frac{x_c^2}{k} + X_{M} +N-2N_R-Mk \Bigg) \numberthis
\end{align*}

Finally, this gives us an expression for the influence gap $G_R$ in favour of the red party for a general set of $l$ isolated cliques, under any party assignment, noting only that $N = N_B+N_R = lk$.
\begin{ceqn}
\begin{equation}
 G_R = \frac{X_{M'}}{N_R}-\frac{X_{M}}{N_B} + \frac{N_B-N_R+Mk}{N_B} -1 + \frac{N_B-N_R}{N_RN_B} \sum_{c=1}^l\frac{x_c^2}{k} \label{eq:IG_caveman_all}
\end{equation}
\end{ceqn}

\paragraph{Influence gap with equal representation} We now show, using the expressions above, what happens in caveman graphs where parties have equal representation, i.e., $N_B = N_R = \sfrac{N}{2}$.
When this is the case a number of terms in Equation \ref{eq:IG_caveman_all} vanish, including the quadratic term, leaving only a linear equation in purely community-level quantities. In particular, note that the term $X_{M'}-X_{M} = (M'-M)\cdot\sfrac{k}{2}$ since all the cliques between $M$ and $M'$ are marginals. After some simplification this gives us an equation for $G_R$ in terms of the number of cliques that are (strictly) dominated by red -- as opposed to individual node quantities. 
\begin{ceqn}
\begin{align*}
 G_R &= \frac{2}{N}(X_M-X_{M'}) + \frac{2M'k}{lk}-1 \\
 &= \frac{2}{lk} (M-M') \frac{k}{2} + \frac{2M'}{l}-1 \\
 &= \frac{M + M'}{l}-1 \numberthis \label{eq:IG_caveman}
\end{align*}
\end{ceqn}

A corollary of this is when $G_R=G_B=0$, when by the above equation 
caveman graphs with equal representation must also have as many strict red majorities as strict blue ones. 

\paragraph{Influence gap and initial majority}

We now turn our attention on how predictive the influence gap is in these structures. We start with the restricted class of caveman graphs with equal representation, which we analysed above.

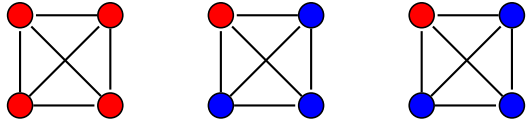
\begin{figure}[ht!]
 \begin{center}
 \begin{tikzpicture}
	[->,shorten >=1pt,auto,node distance=1.2cm,
	semithick]
	\node[shape=circle,draw=black, fill=red] (A) {};
	\node[shape=circle,draw=black, fill=red, right of=A] (B) {};
	\node[shape=circle,draw=black, fill=red, below of=A] (C) {};
	\node[shape=circle,draw=black, fill=red, below of=B] (D) {};
	
	\draw [thick,-] (B) to (A) ;
	\draw [thick,-] (C) to (A) ;
	\draw [thick,-] (D) to (A) ;
	\draw [thick,-] (C) to (B) ;
	\draw [thick,-] (C) to (D) ;
	\draw [thick,-] (B) to (D) ;

	\end{tikzpicture}\hspace{1cm}
	\begin{tikzpicture}
	[->,shorten >=1pt,auto,node distance=1.2cm,
	semithick]
	\node[shape=circle,draw=black, fill=red] (A) {};
	\node[shape=circle,draw=black, fill=blue, right of=A] (B) {};
	\node[shape=circle,draw=black, fill=blue, below of=A] (C) {};
	\node[shape=circle,draw=black, fill=blue, below of=B] (D) {};
	
	\draw [thick,-] (B) to (A) ;
	\draw [thick,-] (C) to (A) ;
	\draw [thick,-] (D) to (A) ;
	\draw [thick,-] (C) to (B) ;
	\draw [thick,-] (C) to (D) ;
	\draw [thick,-] (B) to (D) ;

	\end{tikzpicture}\hspace{1cm}
\begin{tikzpicture}
	[->,shorten >=1pt,auto,node distance=1.2cm,
	semithick]
	\node[shape=circle,draw=black, fill=red] (A) {};
	\node[shape=circle,draw=black, fill=blue, right of=A] (B) {};
	\node[shape=circle,draw=black, fill=blue, below of=A] (C) {};
	\node[shape=circle,draw=black, fill=blue, below of=B] (D) {};
	
	\draw [thick,-] (B) to (A) ;
	\draw [thick,-] (C) to (A) ;
	\draw [thick,-] (D) to (A) ;
	\draw [thick,-] (C) to (B) ;
	\draw [thick,-] (C) to (D) ;
	\draw [thick,-] (B) to (D) ;

	\end{tikzpicture}
 	\caption{A caveman graph with equal representation and IG of $G_R=-1/3$. Notice that red has one strict majority while blue has two, i.e., $M' = M = 1$.}
 	\label{fig:ig0-clique}
 \end{center}
\end{figure}

To do so, we first consider a simple opinion diffusion model, e.g., at each time step
 voters stick to their own opinion unless most of their friends think differently, in which case they flip with probability $\Delta_n$. In Figure \ref{fig:ig0-clique}, for example, we expect such an opinion diffusion model to converge to an outcome where the red party conquers the clique on the left but loses the other two. We note that in this case IG seems to be a good predictor of the overall outcome.
 
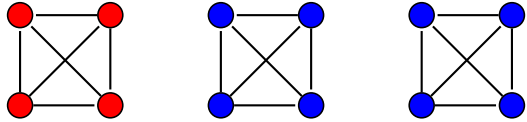
\begin{figure}[ht!]
 \begin{center}
 \begin{tikzpicture}
	[->,shorten >=1pt,auto,node distance=1.2cm,
	semithick]
	\node[shape=circle,draw=black, fill=red] (A) {};
	\node[shape=circle,draw=black, fill=red, right of=A] (B) {};
	\node[shape=circle,draw=black, fill=red, below of=A] (C) {};
	\node[shape=circle,draw=black, fill=red, below of=B] (D) {};
	
	\draw [thick,-] (B) to (A) ;
	\draw [thick,-] (C) to (A) ;
	\draw [thick,-] (D) to (A) ;
	\draw [thick,-] (C) to (B) ;
	\draw [thick,-] (C) to (D) ;
	\draw [thick,-] (B) to (D) ;

	\end{tikzpicture}\hspace{1cm}
	\begin{tikzpicture}
	[->,shorten >=1pt,auto,node distance=1.2cm,
	semithick]
	\node[shape=circle,draw=black, fill=blue] (A) {};
	\node[shape=circle,draw=black, fill=blue, right of=A] (B) {};
	\node[shape=circle,draw=black, fill=blue, below of=A] (C) {};
	\node[shape=circle,draw=black, fill=blue, below of=B] (D) {};
	
	\draw [thick,-] (B) to (A) ;
	\draw [thick,-] (C) to (A) ;
	\draw [thick,-] (D) to (A) ;
	\draw [thick,-] (C) to (B) ;
	\draw [thick,-] (C) to (D) ;
	\draw [thick,-] (B) to (D) ;

	\end{tikzpicture}\hspace{1cm}
\begin{tikzpicture}
	[->,shorten >=1pt,auto,node distance=1.2cm,
	semithick]
	\node[shape=circle,draw=black, fill=blue] (A) {};
	\node[shape=circle,draw=black, fill=blue, right of=A] (B) {};
	\node[shape=circle,draw=black, fill=blue, below of=A] (C) {};
	\node[shape=circle,draw=black, fill=blue, below of=B] (D) {};
	
	\draw [thick,-] (B) to (A) ;
	\draw [thick,-] (C) to (A) ;
	\draw [thick,-] (D) to (A) ;
	\draw [thick,-] (C) to (B) ;
	\draw [thick,-] (C) to (D) ;
	\draw [thick,-] (B) to (D) ;

	\end{tikzpicture}
 	\caption{A caveman graph with $G_R=G_B=0$ and different majorities. Notice how the IG does not ``see" a win for the blue party.}
 	\label{fig:ig1-clique}
 \end{center}
\end{figure}

However, when equal representation is relaxed, IG can also fail to predict robust configurations with clear winners while majority does, as illustrated in Figure \ref{fig:ig1-clique}.

\subsubsection{Towards more realistic communities}

We want to establish how good of a predictor the IG is in graphs with community structure. While we observed interesting edge cases in the simple caveman graphs, it is now important to extend these to more capture more realistic ones.

We do so by generalising them in two ways:

\begin{itemize}
 \item Allowing some edges to be rewired with a set probability, i.e., relaxed-caveman graphs~\cite{Fortunato_2010}.
 
 \item Allowing such probability to depend on homophily, i.e., how likely like-minded voters are to be connected to one another.
\end{itemize}

A {\em relaxed-caveman} graph is a modified version of the basic caveman graph, whereby edges are rewired with some given probability~\cite{Fortunato_2010}. Concretely, given a probability $p_0$ and iterating over all edges $E$ of a set of isolated cliques, an edge $(u,v) \in E$ is rewired as $(u,n)$, for some randomly selected $n \in V$; if $(u,n)$ already exists, nothing happens, such that all new edges are between nodes of different cliques. This extension provides a fairly diverse and intuitively clear set of communities, without the need to rigorously define the concept of community itself or to delve into the plethora of community detection \cite{Fortunato_2010,Schaub2017,Aldecoa2013} and generation \cite{HOLLAND1983109,Lee_2019} methods.

It is important to note that, at low rewire probabilities, the results above on caveman graphs can be extended to relaxed-caveman graphs (and their homophilic variant, see below) by considering the effects of a small number of rewired edges as perturbations $\mathcal{O}(\min(N_R,N_B)^{-1})$, since at worst for a single rewire the assortment of a node gets changed by $\pm 1$ and thus its contribution to the influence gap changes by $\mathcal{O}(N_R^{-1})$ or $\mathcal{O}(N_B^{-1})$. 

The next section will present the model of homophilic relaxed-caveman graphs, where rewiring is linked with homophily, and use it to compare the predictive power of a number of metrics, including the influence gap.

\section{Homophilic Relaxed-Caveman Graphs} \label{sec:RC}

Relaxed-caveman graphs rewire the edges of the original caveman graph {\em without looking at party assignment}. This means that, effectively, the resulting graph, though exhibiting a rich community structure, abstracts away from the relation between connections and opinions, unlike real-world networks, where the two are highly intertwined \cite{Bis09,Bakshy1130}. To address this issue, we propose a modification to the relaxed-caveman graph as the {\em homophilic relaxed-caveman} (hRC) graph model in a similar fashion to the homophilic Erd\H{o}s-R\'{e}nyi and Barab\'{a}si-Albert graphs used in \citet{Tsang-Larson-2016}. This allows us to generate synthetic graphs with communities where the graph structure is dependent on the party assignment, following the observed behavior~\cite{Bis09} that people tend to cluster with people who share their views.
\begin{center}
 \fbox{%
 \parbox{0.7\linewidth}{%
 \textbf{Algorithm: } Homophilic Relaxed-Caveman Graph, $\mathcal{G}$
 \begin{enumerate}
 \item \textbf{Initialise} $G$ as a set of $l$ cliques each of size $k$
 \item \textbf{for} $(u,v) \in E$:
 \begin{itemize}
 \item Choose at random an $n \in V, n \neq v$
 \item \textbf{if} $p(u)$ \textbf{=} $p(n)$; $\tilde{p} = p_0 h$
 \item \textbf{else}; $\tilde{p} = p_0 (1-h)$
 \item Rewire $(u,v)$ as $(u,n)$ with probability $\tilde{p}$
 \end{itemize}
 \item \textbf{return} $\mathcal{G}$ 
 \end{enumerate} 
 }%
}
\end{center}

The algorithm above describes how to generate the hRC graph, starting from a set of disconnected cliques $G=(V,E)$. It becomes hRC graph $\mathcal{G}_p(l,k,p_0,h)$, with $l$ communities, each of size $k$, with {\em rewiring probability} $p_0$ and {\em homophily factor} $h$, given a party assignment $p:V\rightarrow \mathcal{P}$. The probability to rewire $p_0$ can be thought of as the likelihood of changing a pre-existing friendship to a new one, while the homophily is the probability of agent $u$'s new friend $n$ voting for the same party.

We highlight two important subclasses of the hRC model:
\begin{itemize}
\item For $h = 0.5$ we recover the relaxed-caveman graph with rewire probability $p = \sfrac{p_0}{2}$ since a node is equally likely to be rewired to their party as they are to the opposing party, but with probability $p_0 \cdot 0.5$.

\item For high values of $p_0$ nodes from different cliques intermingle sufficiently enough that the community structure begins to fade. 
Community detection methods \cite{Fortunato_2010,Schaub2017}, such as modularity-based methods \cite{PhysRevE.69.066133}, may be used to reestablish communities in this case.
\end{itemize}

\begin{figure}[ht!]
 \begin{center}
 \includegraphics[width=0.9\linewidth]{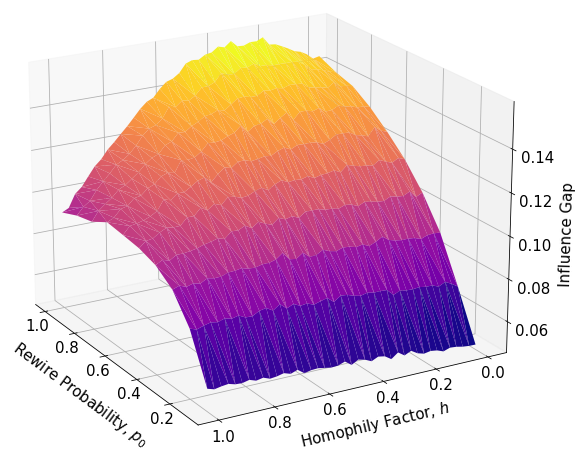}
 	\caption{A surface plot of the mean absolute value of the influence gap, across different homophily factors $h$ and a range of rewire probabilities $p_0$. Each point is measured from $10^4$ party assignments each generating a single graph.}
 	\label{fig:hrc-surf-precise}
 \end{center}
 \vspace{-3mm}
\end{figure}

We now show how the influence gap is distributed in homophilic relaxed-caveman graphs with equal partisan split ($N_R=N_B$). To do so we generated hRC graphs across the entire range of homophily and rewire probability. Due to the randomized nature of the model, for any given set of parameters $h$ and $p_0$, we produced 10,000 graphs and found the mean of their influence gaps, towards whichever party had the higher influence assortment. Note how this, therefore, represents the ability for either party to open an advantage over its opponent, not for a specific one. If, instead, we measured the gap towards a specific party, say red, then we would expect the mean of the gap to be 0 due to symmetry.
We plotted the results in Figure~\ref{fig:hrc-surf-precise}.

We find that for values of $h<0.85$, the absolute value of the influence gap increases monotonically with the rewire probability $p_0$. That is, taking a cross-section for some $h<0.85$ of Figure~\ref{fig:hrc-surf-precise} is monotonic in $p_0$. At higher homophily we find the curves are peaked at around $p_0 \approx 0.7$.

We see a more complex phenomenon when taking cross-sections in $p_0$, that is, curves change from nearly flat at low $p_0$ to being peaked at higher $p_0$. Specifically, the influence gap is asymmetrically unimodal, with a peak at around $p_0 = 1$ and $h = 0.3$. Notice that this means parties are marginally better off when the likelihood of forming a community between the party members is lower than with the opposing party. This asymmetry, we believe, is in large part due to the definition of the influence assortment for a node. When the neighbourhood $\mathcal{N}(n)$ of an agent $n$ is just slightly in support of the opposite party but their own preference is enough to push the poll into a (weak) majority $\Delta_n \geq 0.5$ then the maximal value of the influence gap is reached, when most/all nodes face this situation.

\section{Predicting Two-Party Elections: The Model}\label{sec:dynamics}

\begin{figure}[ht!]
 \begin{center}
 \includegraphics[width=0.9\linewidth]{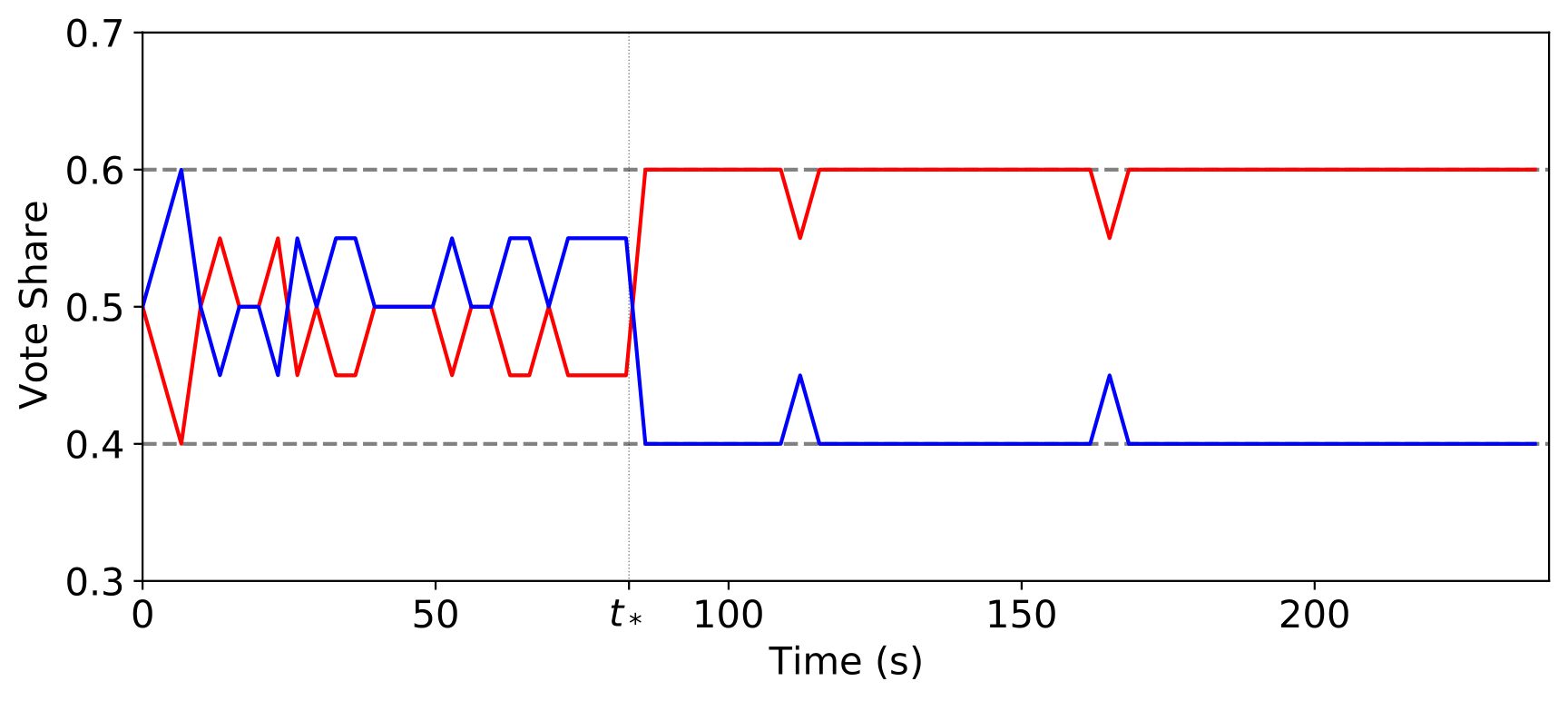}
 	\caption{Example time series of a voting game, using \citet{Stewart2019} behavioural model on a homophilic relaxed-caveman graph with rewire parameter $p_0 = 0.3$ and homophily $h = 0.3$. The two parties initially have equal vote share but eventually the red party reaches a super-majority of $V = 0.6$ (the dashed lines are $V$ and $1-V$). The vertical dotted line at $t_*$ represents the transition between early and late phases of the game.}
 	\label{fig:dynamics_example}
 \end{center}
 \vspace{-3mm}
\end{figure}

In this section we examine whether the findings of \citet{Stewart2019} still hold for two-party elections when communities are present. Our conclusion, in summary, is that influence gap does not predict influence dynamics when initial votes are equally split, which is a central assumption in their analysis. When this assumption is relaxed, the correlation is restored, but a more predictive and computationally simpler metric exists, namely counting the initial majority of either party. This holds across all levels of homophily and rewiring, with some interesting specificities.

 Next, we provide the details of our experiment and results, which uses the same behavioural parameters of \citet{Stewart2019}, starting from the voter model.

\subsection{Voter Model}
For $N$ voters, at least half are assigned to the red party and the remainder to the blue, and all are placed in an influence network, in line with~\citet{Stewart2019}. A voter's knowledge is restricted by their social network, only knowing the voting intentions of their neighbours as well as their own, serving as a form of poll, to which they wish to conform. The game lasts for a fixed amount of time, during which players can change their voting intentions synchronously\footnote{Unlike the iterative voting model of \citet{Tsang-Larson-2016}, in which every voter updates separately to the others, this is more akin to the synchronous updates of \citet{Alon-Feldman-Omer-2015}.}. The winning party is the one to hold a majority above a threshold $V > 0.5$ when the updating process is done, otherwise we consider it a deadlock.

The agents follow the stochastic behavioural model developed by~\citet{Stewart2019}, informed by a social experiment with human subjects who were given pay-offs depending on the success of their assigned party. At any given time, a voter, according to the behavioural model, would vote for their assigned party with a probability that depends on: a) what their surroundings predict will happen (win, lose or deadlock) and b) the stage of the game (early or late). In other words, for each individual there exists a family of six parameters $p_{ij}$, where $i \in \{$win, lose, deadlock$\}$ is the poll's prediction and $j \in \{$early, late$\}$ is the stage of the game, that are precisely these probabilities, henceforth \textit{strategies}.

\begin{table}[htb]
\centering
\begin{tabular}{ c|c c} 
 \hline
 $\bar{p}_{ij}$ & \textit{Early} & \textit{Late} \\
 \hline
 \textit{Win} & 0.975 & 0.979\\
 \textit{Deadlock} & 0.964 & 0.911\\
 \textit{Lose} & 0.598 & 0.574\\
 \hline
\end{tabular}
\caption{Mean agent strategies $\bar{p}_{ij}$, such that an agent with a neighbourhood poll predicting state $i$ during phase $j$ will stick to their party, on average, with probability $\bar{p}_{ij}$. These values were inferred from social experiments (see the Supplementary Material of \citet{Stewart2019} for an extensive discussion of how these numbers are obtained and why these are used independently of the graph structure).}
\label{table:pij_means}
\end{table}

Since voters are not a homogeneous bunch and will have different strategies, each parameter is sampled from the empirical distribution of the social experiment. Thus while $p_{ij}$ is a random variable, which has an empirical distribution with mean given by Table \ref{table:pij_means}, for each voter $v$ a set of 6 parameters $p_{ij}^v$ are realisations of the random variables.
We point out that the behavioural parameters can be used independently of the initial structure, as voters are unaware of the graph they sit in, an assumption also made by \citet{Stewart2019} when replicating their findings with simulations.

\subsection{Benchmark Metrics} \label{sub:benchmark}
 
Since an exact analysis of the opinion dynamics faces significant complexity barriers, \citet{Stewart2019} proposed using the Influence Gap as a prediction tool of the dynamic's outcome. To the IG proposal we add three more metrics, which we will compare against:

\begin{description}
 \item [Majority] The initial majority of the red party. 
 \item [Deterministic voter skew (dVS)] A deterministic simplification of the update dynamics; at each time step every agent synchronously conforms to the strict majority party in their poll, keeping the current choice in case of a tie, and after $\sigma$ steps the voter skew is measured. In principle, one could evolve the system for as many steps $\sigma$ as in the stochastic process, but we use $\sigma=1$, as errors due to the simplification may be propagated and worsened with more steps.
 \item [Efficiency gap (EG)] A classical political science metric~\cite{efficiencygap}, developed to measure gerrymandering in two-party elections, in which we examine how many votes were ``wasted'', i.e., could have been eliminated without changing the outcome.
\end{description}

\subsection{Experimental Setup}

We have $N = 20$ voters, each assigned a party and then are placed in a graph $G$, generated by a number of different assignments of $h$ and $p_0$. Following the setting of \citet{Stewart2019} (which was also based on experiments with people), each simulation game runs for $240$ seconds, which starts with the early phase for $83$ seconds, and then transitions to the late phase. During the game, every $3.3$ seconds a voter, $v$, can update their voting intention, with probability $p_{ij}^v$, which are sampled from the empirical parameter-distributions found from the human social experiment~\cite{Stewart2019}. In total, for a single simulation $N$ samples are taken from $6$ distributions each (for each stage -- early/late -- and each voting outcome -- Red victory/Blue victory/Deadlock). After $240$ seconds have elapsed the vote share across the entire graph is measured.


An example of a time series produced by a simulation is shown in Figure~\ref{fig:dynamics_example}. The different sections of the plot, partitioned by dashed and dotted lines, represent different strategies. For example, in the early phase ($t<t_*$) both parties are deadlocked and thus agents vote for their assigned party with probability $p_{\text{deadlock},\text{early}}^{\text{agent}}$. The convergence of a time series is not a guaranteed because, as in the original social experiment, the game finishes after $240$ seconds.

We simulate over $10^4$ elections for each set of hRC parameter values $(p_0,h)$ while varying the initial number of red nodes, $N_R\geq 10$. Parameters were chosen to cover a reasonable range, $p_0 \in \{0,0.4,1\}$ and $h\in \{0,0.1,\cdots,1\}$, so that in total there were over $33,000$ simulated elections. At the start of each election we measure the influence gap, the deterministic voter skew and the efficiency gap (see Section \ref{sub:benchmark}); the final outcome of the election is also recorded as a vote skew towards the red party $\sfrac{N_R}{N}-\sfrac{1}{2}$. For example, Figure~\ref{fig:ig+maj_vs} shows the elections occurring on hRCs with $(p_0,h)=(0.4,0.6)$ and illustrates how two metrics, the IG and majority, both correlate strongly with the final outcome (note that the figure shows a case where parties did not have equal power initially).

\begin{figure}[t!]
 \centering
 \begin{subfigure}{0.49\linewidth}
 \centering
 \includegraphics[width=1\linewidth]{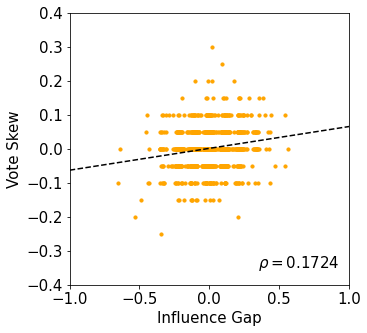}
 \label{fig:dynamics-RC}
 \end{subfigure}%
 \begin{subfigure}{0.49\linewidth}
 \centering
 \includegraphics[width=1\linewidth]{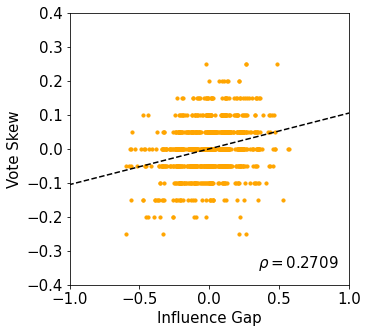}
 \label{fig:dynamics-hRC}
 \end{subfigure}
 \caption{Simulations of the behavioural model of \citet{Stewart2019} acting on relaxed- and homophilic relaxed-caveman graph, starting with equal representation. In both cases the influence gap correlates very weakly with voter skew, having a Pearson's $\rho < 0.3$, and passing a significance test with $p < 10^{-7}$. \textbf{Left: } relaxed-caveman graphs with rewire parameter $p=0.3$. \textbf{Right: } homophilic relaxed-caveman with rewire probability $p_0 = 1$ and homophily factor $h = 0.3$.} \label{fig:single_maj_corr}
 \vspace{-3mm}
 \label{fig:equal-representation}
\end{figure}

\section{Predicting Two-Party Elections: Results} \label{sub:res}
 Where \citet{Stewart2019} found strong correlation between IG and election outcome in scale-free (Barab\'{a}si-Albert) graphs for a given initial voter skew, we find the contrary in RC and hRC graphs. Starting with equal representation -- same number of red nodes as blue -- we find that the presence of communities suppresses the correlation noted there. In both the relaxed-caveman and the homophilic relaxed-caveman with given parameter sets, the Pearson correlation coefficient is small -- $\rho < 0.3$ -- as seen in Figure \ref{fig:single_maj_corr}.

\begin{figure}[!h]
 \centering
 \includegraphics[width=0.9 \linewidth]{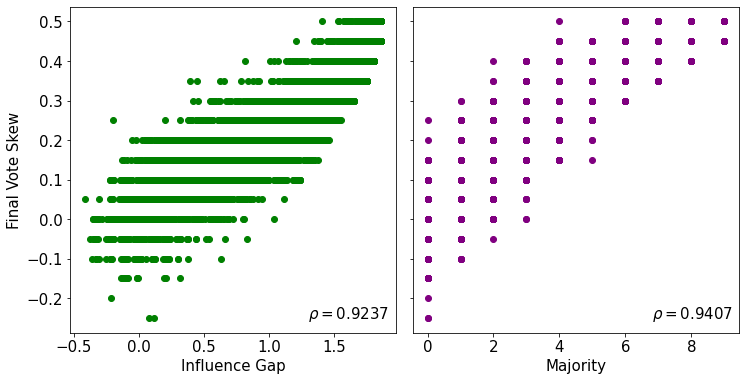}
 \caption{For a hRC with $p_0 = 0.4$ and $h=0.6$, we plot the final vote skew against the IG and the majority of the red party. }
 \label{fig:ig+maj_vs}
\end{figure}

When we allow for unequal initial setup, results show more complexity.
In Figure \ref{fig:ig+maj_vs} we plot as an example how the initial influence gap and majority compare to the final election outcome, for a hRC graph with intermediate rewire $p_0 = 0.4$ and intermediate homophily $h = 0.6$. We see in both cases very strong correlations with a high Pearson correlation coefficient (PCC) $\rho > 0.9$. Note that by symmetry near identical distributions would be found, if we had simulated $N_R<10$.

\begin{figure*}[!tb]
 \centering
 \includegraphics[width=0.9\linewidth]{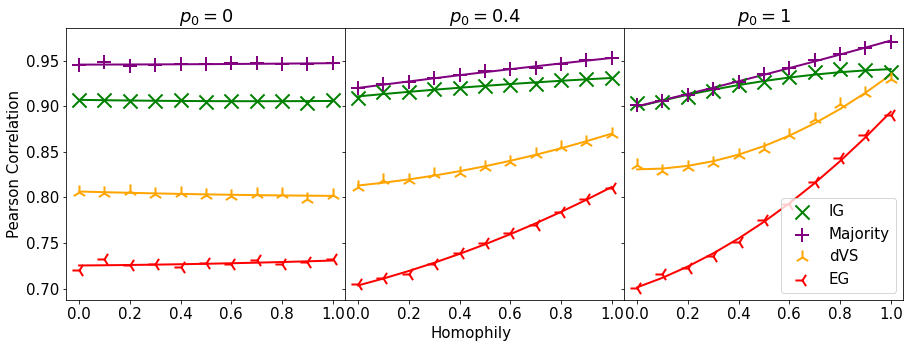}
 \caption{As model parameters of the homophilic relaxed-caveman are varied independently, the Pearson correlation coefficient (PCC) between the final voter skew and the majority (purple), the influence gap (IG, green), the deterministic voter skew (dVS, orange) and the efficiency gap (EG, red) are plotted. Quadratic lines of best fit for multiple rewire probabilities are plotted, on the left $p_0 = 0$, in the middle $p_0 = 0.4$ and on the right $p_0 = 1$. The influence gap only outperforms the majority once, at $(p_0,h)=(1,0)$.}
 \label{fig:pcc_lines}
\vspace {-1mm}
\end{figure*}

In other words, for this set of parameter values, both IG and majority are strong predictors of the final outcome -- in so far as their initial values correlate strongly with the result. However we can begin to see how the influence gap does not predict the outcome as accurately as the initial majority.
{\subsection{The Effects of Rewiring and Homophily}}
We now present our findings taking rewiring and homophily into account, a snapshot of which is shown in Figure \ref{fig:pcc_lines}. 

Before delving into discussing them, it is important to note that neither homophily nor rewiring correlate, on their own, with the final outcome of the election, with Pearson correlation coefficients of $\rho = -0.0256$ and $\rho=0.0106$ respectively. This does not imply that neither have an effect at all -- indeed, Figure \ref{fig:pcc_lines} shows otherwise -- simply that they do not impact the election single-handedly. 


\paragraph{High Homophily} When $h>0.5$, hRC graphs are rife with echo chambers in which an agent has friends mostly of the same party and opinion as themselves. Their polls are therefore fairly homogeneous and they see no compelling reason to change their vote, meaning the graph structure plays little to no role. Extending this intuition to all agents in the graph, very little diffusion of opinions occurs and thus the final outcome of the election will not be too dissimilar to the starting one. As such, counting the number of votes at the start will closely resemble -- and thus predict -- the number of votes at the end, and majority here can better predict the outcome than IG. Moreover, this logic can help explain the general trend that the correlation of most metrics increase drastically at high homophily.

More formally, at high $h$ the average poll will show a majority towards their agent's party $\Delta_v > 0.5$ or even a super-majority $\Delta_v > V$. During phase $j$ of the voting game, most voters $v$ see a prediction of $i = \text{win}$, so that their strategy is $p_{\text{win},j}^v$. Since the empirical distribution for $p_{\text{win},j}$ is heavily biased towards $p_{\text{win},j} \geq 0.9$ most voters will stick to their initial opinion. 

\paragraph{Low Homophily} Conversely, for $h < 0.5$, an agent's poll, at least initially, likely shows a super-minority $\Delta_v < 1-V$ or equivalently $i = \text{lose}$ such that over 40\% of agents will change their votes. In other words, most polls are quite diverse and an agent will see mostly nodes of the opposite party and ``doubt" their opinion more often.

In this scenario the dynamics become less stable, and more complex diffusion occurs. Therefore, most metrics perform either as well as or worse at $p_0=1$ than $p_0=0$. Only the deterministic voter skew sees an increase in predictive power as $p_0$ increases, at all values of homophily. We suspect this is as a single deterministic step causes the network to evolve towards a more stable configuration and, as such, closer to the final outcome.

Another effect of this is that the predictive powers of both majority and influence gap monotonically both decrease as homophily decreases; in fact for $h < 0.5$ the difference in PCC between IG and majority shrinks considerably.

As we move up the rewiring level, the majority PCC is linear while the IG PCC is a flat quadratic. Both the efficiency gap and deterministic voter skew, however, show a move from linear to quadratic as communities become more diluted. Moreover, both metrics are consistently outperformed by influence gap, though still have reasonably high correlations; once again majority bests all others.

Overall, only once does the influence gap predict outcomes better than the initial majority, at $p_0=1$ and $h=0$, in other words, when nodes are extremely diverse and seek out alternative opinions to their own. The difference, however, is tiny at $|\Delta\rho| = 0.00144$ -- and likely due to the stochastic nature of the elections. We conclude that simply counting the number of party votes is a better predictor than the metric of \citet{Stewart2019}.

\subsection{Regression Models} \label{sub:reg_models}\label{sec:comb_met}
In order to further explore the role of majority and influence gap ($x_M$ and $x_G$ respectively, following typical regression notation) as predictive or explanatory variables we use linear regression (Equation~\ref{eq:lin_reg}) to build several models of the vote skew, $y$. Two models are single-featured using only the initial majority or the influence gap and the third is a joint model built using a multiple regression of both features. All three are trained on the same $70\%$ of the data and tested on the remaining $30\%$. 
\begin{equation}
 y = \beta_M x_M + \beta_G x_G + \beta_0 \label{eq:lin_reg}
\end{equation} 

\begin{table}[htb]
\centering
\caption{Coefficients of regression, $\beta$, and of determination, $R^2$, for regression models of the dynamic voting game outcome on homophilic relaxed-caveman graphs for two-party elections.} \label{table:lin_reg}
\begin{tabular}{ |c|c c c|c| } 
 \hline
 \textit{Metric} & \textit{$\beta_M$} & \textit{$\beta_G$} & \textit{$\beta_0$} & \textit{$R^2$}\\ \hline
 Majority & 0.0507 & & 0.0363 & 0.881\\
 IG & & 0.226 & 0.00474 & 0.837\\
 Majority, IG & 0.0333 & 0.0869 & 0.0144 & \textbf{0.902}\\
 \hline
\end{tabular}
\end{table}

The regression confirms our observations , that majority is a better predictive tool than influence gap. This does not, however, render IG useless. In particular the joint model outperforms both individual models that either use majority or influence gap exclusively as shown in Table \ref{table:lin_reg}, despite some colinearity between features. In other words, IG is an informative metric that can build upon the simple predictions of majority, but alone is not as effective.

\section{Extending the Influence Gap to Multiple Parties} \label{sec:multiparty}
In many political systems more than two parties are competitive and relevant. While the US has two main parties, Canada has 5 parties with more than 5\% in parliament; the UK has 3 such parties, and the situation is similar in most Continental European democracies and around the world. In this section we will discuss the challenges in expanding the definitions for IG to the multi-party setting, by showing a few possible definitions for the influence assortment and gap. We then apply these definitions to small graphs to highlight the subtle differences and to motivate the choice of one definition over the other.

\subsection{Influence Assortment}
We examine two possible extensions to multi-party settings. Both assume, naturally, that plurality is the reasonable extension to majority (which always exists in two-party settings, but not guaranteed to exist with more parties). The first assumes the ``force'' of the plurality winner is all there is, and thus even when an agent does not support it, the negative weight it carries is only as much as the plurality winner could muster.

\begin{equation}
 a_n = \begin{cases}
 \Delta^{p(n)}_n & \text{if } \Delta^{p(n)}_n = \max(\bm{\Delta}_n), \\
 -\max(\bm{\Delta}_n) & \text{otherwise.}
 \end{cases} \label{eq:a_n1}
\end{equation}

An alternative definition takes an all-against-one stance: rather than taking the most dominant party, we sum over all other parties. That is, when an agent does not support the plurality winner, their negative weight is the sum of all the parties one did not support. 

\begin{equation}
 a_n = \begin{cases} 
 \displaystyle \Delta^{p(n)}_n & \text{if } \Delta^{p(n)}_n = \max(\bm{\Delta}_n), \\
 -(1-\Delta^{p(n)}_n) & \text{otherwise.}
 \end{cases} \label{eq:a_n2}
\end{equation}

Both definitions reduce to the original two-party definition of influence assortment and have reasonable intuition behind them. 

\subsection{Influence Gap}
There are two main ways to define the influence gap. First, to define it as the difference between influence assortments of party $P$ and the (next-) most \textit{influential} party. Second, as the difference between party $P$ and the (next-) plurality party (i.e., plurality runner-up).

\begin{align}
 G_P &= A_P - \max_{Q\neq P}A_Q \label{eq:G_P1}\\
 G_P &= A_P - A_Q \quad \text{where } Q = \argmax_{P'\neq P} N_{P'} \label{eq:G_P2} 
\end{align}

These two gaps largely correlate with one another, since influence and partisan split are well correlated -- though not without exceptions. Specifically, the latter definition can produce non-unique values for the same network.\\

To more carefully explore our suggested extensions, we will now examine several settings of multi-party situations, and for that we shall explore the case of three parties: $\mathcal{P} = \{\circ, \triangle,\square\}$.

\subsection{$N=3$ Examples}
Consider three voters who each vote for a different party. There are only two connected networks to consider: the cycle and the line. Due to the symmetry of the cycle, all influence assortments are equal and thus all influence gaps are 0. For the line where $\triangle$ is the middle node, there is symmetry between the $\circ$ and the $\square$.

\begin{table}[!htbp]
\centering
\caption{Comparison of definitions, applied on $N=3$ graphs.}
\begin{tabular}{*6c}
\hline
Graph & $a_n$ & $G_P$ & $G_\circ$ & $G_\triangle$ & $G_\square$\\
\hline
Cycle & (*) & (*) & 0 & 0 & 0 \\

\hline
\multirow{2}{*}{$\circ-\triangle-\square$} & \multirow{2}{*}{(*)} & (\ref{eq:G_P1}) & 0&-\sfrac{1}{6}&0 \\
{}& {}& (\ref{eq:G_P2}) & 0, \sfrac{1}{6}&-\sfrac{1}{6}&0, \sfrac{1}{6} \\
\hline
\end{tabular}
\end{table}

For the line graph, notice that the definition of assortment leads to $\triangle$ being comparatively heavily influenced (Equation~\ref{eq:a_n2}), or only slightly influenced (Equation~\ref{eq:a_n1}). Moreover, the definition of the influence gap can lead to multiple results for the same graph -- this is clearly problematic. In general the latter definition of IG, Equation~\ref{eq:G_P2}, produces non-unique results if there are joint winners -- this can occur frequently and as such we hence only consider the first definition, Equation~\ref{eq:G_P1}.

\subsubsection{$N=4$ Examples} \label{sub:N4}
Consider a network of 4 voters, two of which vote for the $\circ$ party, one votes $\triangle$ and another votes $\square$. Already with this set-up there at least a dozen configurations to create a connected network; here we consider a few of these to illustrate the simple examples that complicate choosing a definition, starting with the 4 non-trivial line graphs (since swapping $\square$ and $\triangle$ produces a qualitatively identical result).

\begin{table}[!htbp]
\centering
\caption{Comparison of definitions, applied on lines of size $N=4$.}
\begin{tabular}{*5c}
\hline
Line & $a_n$ & $G_\circ$ & $G_\triangle$ & $G_\square$\\
\hline
{$\circ-\circ-\triangle-\square$} & (*) & \sfrac{1}{3}&-\sfrac{1}{2}&-\sfrac{1}{3} \\

\hline
{$\circ-\triangle-\square-\circ$} & (*) & \sfrac{1}{6}&-\sfrac{1}{6}&-\sfrac{1}{6} \\

\hline
{$\triangle-\circ-\square-\circ$} & (*) & -\sfrac{1}{12}&\sfrac{1}{12}&-\sfrac{13}{12} \\

\hline
$\triangle-\circ-\circ-\square$ & (*) & \sfrac{1}{6}&-\sfrac{1}{6}&-\sfrac{1}{6} \\
\hline
\end{tabular}
\end{table}

We see that for such line graphs, both assortment definitions coincide. In general, for any sized line graph (or regular graph of degree 2), definitions (\ref{eq:a_n1}) and (\ref{eq:a_n2}) always coincide. For any number of parties in $\mathcal{P}$, a node either has one neighbour or two. When $n$ has exactly one neighbour -- in other words, $n$ is at the end of the line -- $\Delta^{p(n)}_n = \sfrac{1}{2}$ or $1$, so that $\Delta^{p(n)}_n = \max(\bm{\Delta}_n)$ hence both definitions coincide. In the bulk of the line, with two neighbours, there are three cases:

\begin{itemize}
 \item $p(n)=p(n-1)$ or $p(n+1)$; $p(n)$ has plurality in the poll of $n$, therefore $(\ref{eq:a_n1}) = (\ref{eq:a_n2})$ trivially.
 \item $p(n-1)=p(n+1)\neq p(n)$; $p(n-1)$ holds plurality but is the only alternative party to $p(n)$, hence $\max (\bm{\Delta}_n) = 1 - \Delta^{p(n)}_n (= \sfrac{2}{3})$.
 \item $p(n-1)=p(n)=p(n+1)$; all three parties hold joint plurality, therefore $(\ref{eq:a_n1}) = (\ref{eq:a_n2})$ trivially.

\end{itemize}

Using similar analysis, ring graphs (with degree 2) also cause the two definitions to coincide.

\begin{table}[!htbp]
\centering
\caption{Comparison of definitions, applied on stars of size $N=4$, with different centres.}
\begin{tabular}{*5c}
\hline
Centre & $a_n$ & $G_\circ$ & $G_\triangle$ & $G_\square$\\
\hline
\multirow{2}{*}{$\triangle$} & (\ref{eq:a_n1}) & 0&-1&0 \\
{} & (\ref{eq:a_n2}) & 0&-\sfrac{5}{4}&0 \\

\hline
$\circ$ & (*) & \sfrac{1}{4}&-\sfrac{1}{4}&-\sfrac{1}{4} \\
\hline
\end{tabular}
\end{table}

On the other hand as soon as a node has degree $3$ or higher with at least 3 parties we start to see non-unique solutions. For example, examine star graphs (i.e., graphs in which one node is connected to all others, and no other edges exist) when there are two $\circ$'s, one $\triangle$ and one $\square$. When the $\triangle$ is at the centre, $\circ$ holds plurality but is not the only alternative party. Hence Definition (\ref{eq:a_n2}) will produce a more negative assortment and thus a more extreme influence gap.

More complex edge configurations exist, but we look at two more to further illustrate the problem: a clique and a near-clique -- i.e., a clique where a single edge ($\circ-\triangle$) has been removed.

\begin{table}[!htbp]
\centering
\caption{Comparison of definitions, applied on near-cliques of size $N=4$, with different centres.}
\begin{tabular}{*5c}
\hline
Graph & $a_n$ & $G_\circ$ & $G_\triangle$ & $G_\square$\\
\hline
\multirow{2}{*}{Near-Clique} & (\ref{eq:a_n1}) & \sfrac{5}{6}&-\sfrac{5}{6}&-\sfrac{5}{3} \\
{} & (\ref{eq:a_n2}) & \sfrac{5}{6}&-\sfrac{5}{6}&-\sfrac{23}{12} \\

\hline
\multirow{2}{*}{Clique} & (\ref{eq:a_n1}) & 1&-1&-1 \\
{} & (\ref{eq:a_n2}) & \sfrac{5}{4}&-\sfrac{5}{4}&-\sfrac{5}{4} \\
\hline
\end{tabular}
\end{table}

Henceforth we consider only Definition (\ref{eq:a_n1}) for influence assortment and Definition (\ref{eq:G_P1}) for influence gap, due to their faithfulness to the original concepts and, in the case of IG, that they are well-defined, i.e., do not produce multiple results on the same graph.

\section{Influence Gap on Large Graphs with Multiple Parties}\label{sec:3-party-large}
In this section, we use the previously identified extensions of influence gap and assortment, to handle larger social networks with multiple parties. We first provide an analysis of their behaviour on cliques, for any number of parties, and then we show experimental evidence of their predictive power on hRC graphs.
\subsection{Cliques and Plurality Centres}
Consider a clique of size $N$ with a set of parties $\mathcal{P} = \{0,\cdots,|\mathcal{P}|-1\}$ -- in other words, every voter has perfect information on all other voters -- and let $N_P$ be the number of voters for party $P$. In this situation \textit{influence} perfectly correlates with voter split, i.e., the fraction of the population. There are two main cases to consider: 1) a single plurality winner, 2) multiple joint winners (a ``deadlock'' in the parlance of \citet{Stewart2019}).

In case 1, let us denote by $P=0$ the singular winner and note that $\max(\bm{\Delta}_n) = \frac{N_0}{N}, \forall n \in V$ since it is a clique. Moreover, for all voters of a single party $P$, the influence assortment is the same, due to the symmetry of the system, and hence the influence assortment at the party level is equal to the assortment at the node level. From Definition (\ref{eq:a_n1}) we can conclude that $A_0 = \frac{N_0}{N}$ and $A_P=-\frac{N_0}{N}, \forall P\in\mathcal{P}\setminus 0$. As such the influence gap for the plurality winner is $+2\frac{N_0}{N}$, and for the losers $-2\frac{N_0}{N}$.

In case 2, let $W$ be the number of jointly winning parties $\mathcal{W} = \{0,\cdots,W-1\}$, with $N_0$ number of voters. In this case for $P<W$, $A_P=\frac{N_0}{N}$ with the remainder as $A_P=-\frac{N_0}{N}$ for $P\geq W$; the influence gap follows simply as $+2\frac{N_0}{N}$ for the winners and $-2\frac{N_0}{N}$ for the losers. 

Combining the two cases, we find the influence gap for a clique of size $N$ with $W$ (joint) winners ($P<W$), who have won with $N_0$ number of voters each. The magnitude of the gap is the same for all parties, but whose sign is determined by being a winning party or not.

\begin{equation}
 G_P = \begin{cases}
 +\frac{2N_0}{N} & \text{for } P<W,\\
 -\frac{2N_0}{N} & \text{otherwise}
 \end{cases} \label{eq:G_P-complete}
\end{equation}

From a clique we can find the influence gap of other networks by removing edges systematically from it. Consider removing an edge between two nodes $u$ and $v$, neither of which vote for a plurality party, $P_0$. In doing so, in the polls of $u$ and $v$, $P_0$ (and all other plurality winners) still holds plurality and even has a higher poll fraction $\Delta^{P_0}_u=\Delta^{P_0}_v=\frac{N_0}{(N-1)}$. As such $a_u$ and $a_v$ decrease, so that overall $G_{P_0}$ increases while $G_{p(u)}$ and $G_{p(v)}$ decreases.

Repeating this procedure one can keep removing edges until plurality winners form a center or core (of size $W N_0$) -- in other words, all voters of $P<W$ are connected to all other nodes -- while those in the periphery each have $WN_0$ neighbours. Thus all nodes have polls that show plurality towards $P<W$ with fraction $\Delta^{P<W}_n = \frac{1}{W+\frac{1}{N_0}}\approx\frac{1}{W}$ such that the influence gap is simple.

\begin{equation}
 G_P = \begin{cases}
 +\frac{2N_0}{WN_0+1} & \text{for } P<W,\\
 -\frac{2N_0}{WN_0+1} & \text{else.}
 \end{cases} \label{eq:G_P-plurality-core}
\end{equation}

Any network between this plurality-core and the clique has an influence gap bounded by Equations \ref{eq:G_P-complete} and \ref{eq:G_P-plurality-core}: $\frac{2N_0}{WN_0+1}\leq|G_P|\leq\frac{2N_0}{N}$, with the same dependency on winning for the sign.

\subsection{Predicting Three-Party Elections} \label{sec:3-party-hrc}
Equipped with the formal multiparty definition for \textit{influence assortment} (Equation~\ref{eq:a_n1}) and \textit{influence gap} (Equation~\ref{eq:G_P1}), we can replicate the methodology of Section \ref{sec:dynamics} in order to simulate elections with $3$ parties -- red, blue and green. The voter model easily extends to 3 or more parties; with probability $p_{ij}^v$ a voter $v$ in state $i$ in phase $j$ of the election sticks to her current vote, and with probability $1-p_{ij}^v$ she votes for the party with highest poll fraction (excluding her own). Similarly the algorithm for the homophilic relaxed-caveman graph works for any arbitrarily sized set of parties $\mathcal{P}$. 

For the benchmark metrics, we focus only on party votes -- counting the number of voters for each party. For simplicity, we omit the analysis of the simplied dVS dynamic and the two-party intended EG.

\begin{figure*}[h!]
 \centering
 \includegraphics[width=\linewidth]{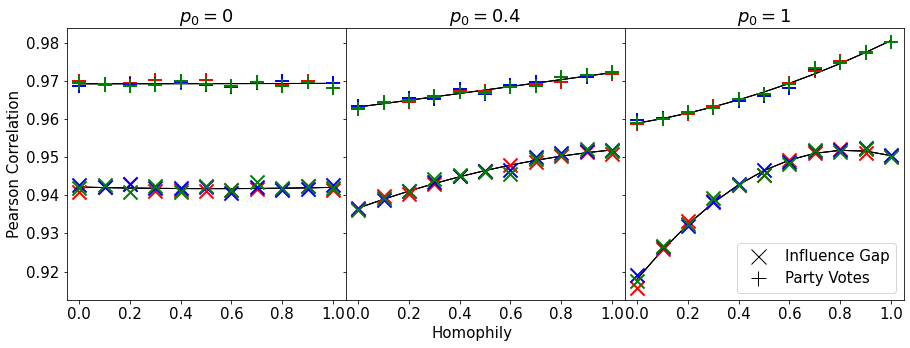}
 \caption{For a three-party election occurring on hRC graphs, the Pearson correlation coefficient (PCC) between the final vote skew and the initial number of party voters (pluses) and influence gap (crosses) are plotted for each party colour (red, blue or green). As before for different values of the rewire -- $p_0 = 0$ on the left, $p_0 = 0.4$ in the middle and $p_0=1$ on the right -- the homophily is varied. In black are lines of best fit for each metric, across all parties. For visual clarity we leave out the deterministic voter skew and efficiency gap, the latter being canonically used for the American two-party system hence unsuitable for the multi-party case.}
 \label{fig:multiparty_pcc_lines}
\vspace {-1mm}
\end{figure*}

Looking at Figure \ref{fig:multiparty_pcc_lines}, there is a very clear and consistent rift between the influence gap and the party votes -- IG is always the weaker predictor. Moreover, while in the two-party case the difference between PCCs can be as small as $|\Delta\rho|\geq 0.00141$, the smallest difference for the three-party case is $|\Delta\rho|\geq0.0191$. In other words, in the three-party case the influence gap never gets arbitrarily close in its predictive accuracy to the simple vote count, there seems to be a bound on how close the IG will get. 

Another curious effect in moving to the three-party case is the behaviour with high rewiring probability. In this case, the PCC for influence gap becomes noticeably worse for low homophily, causing a widening divide between it and the party votes, in direct contrast to the two-party case where the difference shrinks.

As a conclusive remark, we note that for, more than three parties and random initial assignment and rewiring, the share of any winning party in any neighbourhood is smaller in expectation. Therefore, the magnitude of the assortment values $a_n$ (for any node $n\in V$) is smaller, and so is the influence gap, leading it to be an even noisier predictor.

\section{Conclusion}\label{sec:discussion}

We studied the predictive power of a recently proposed metric, the influence gap, to compute the results of many-party elections on networks with community structure. To do so, we proposed a novel model, the homophilic relaxed-caveman, as a means to generate synthetic graphs with communities that may exhibit echo chambers. 
In order to extend our analysis to the multi-party case we then proposed multiple definitions for the influence assortment and gap. Their similarities and subtle differences were illustrated first in small simple graphs and then characterised for large graphs -- such as cliques and plurality centres. 

We showed that the presence of communities suppresses the power of the influence gap as a predictor of the final outcome. A much simpler metric -- the initial majority -- is a far better predictor, albeit one which can be improved somewhat by combining it with the influence gap for an even better predictor. Surprisingly, the initial advantage is an even stronger metric in the multi-party cases.

Having measured the efficacy of several metrics, this poses the question of whether a metric exists that can predict the voting dynamics most accurately, while still being easy to compute. 

A further important direction concerns the difference in voters' behaviour. In our hRC models, the homophily level and the probability of rewiring are the same for both parties. However, we may want to distinguish between electorates that have different levels of approaching others. For example more open-minded voters, who do not mind accepting connections that do not share their view, while others are more close-minded. Preliminary results suggest that a party can leverage a higher homophily factor in this way to structure the network to its favour.


The dynamic model has only shown that the outcome of an election on networks with community can be biased by the structure. However, we may want to model manipulation explicitly, by allowing parties to insert artificial bots or zealots, as \citet{Stewart2019} did, to influence members behaviour. Just like the forceful agents in \citet{acemoglu-spread}, these can be modelled as nodes that have a party affiliation but never update their view. 

Finally, agents may be allowed to actively seek new friendships, whether to express their views more widely or to receive more opinions -- the distinction becoming important when considering directed networks. Can a party utilise these dynamic connections to its advantage?

\section*{Acknowledgement}
This work was supported by ISF grant 1965/20; GIF grant I-2527-407.6/2019; and EPSRC grant EP/S022244/1 for the MathSys CDT.

\bibliography{mybibfile}

\end{document}